\documentclass[reqno,11pt]{amsart}
\usepackage{amscd,amssymb,verbatim}
\numberwithin{equation}{section} \theoremstyle{plain}

\newcommand\alp{\alpha}         
\newcommand\bet{\beta}
\newcommand\gam{\gamma}         \newcommand\Gam{\Gamma}
\newcommand\del{\delta}         
\newcommand\eps{\varepsilon}
\newcommand\zet{\zeta}
\newcommand\tet{\theta}

\newcommand\lam{\lambda}                \newcommand\Lam{\Lambda}
\newcommand\sig{\sigma}

\newcommand\RR{\mathbb{R}}

\newcommand\ZZ{\mathbb{Z}}

\newcommand\CC{\mathbb{C}}

\newcommand\sdp{\times \hskip -0.3em {\raise 0.3ex
\hbox{$\scriptscriptstyle |$}}} 

\newcommand\Det{\operatorname{Det}}

\newcommand\Id{\operatorname {Id}}

\newcommand\IM{\operatorname{Im}}

\newcommand\MOD{\operatorname{mod}}

\newcommand\rank{\operatorname{rank}}

\newcommand\RE{\operatorname{Re}}
\newcommand\Res{\operatorname{Res}}

\newcommand\Tr{\operatorname{Tr}}

\newcommand\olam{{\overline{\lambda}}}

\newcommand\hatr{{\hat{r}}}
\newcommand\hatR{{\widehat{R}}}

\newcommand\tiltet{{\widetilde{\theta}}}

\theoremstyle{plain}
\newtheorem{Thm}[subsection]{Theorem}
\newtheorem{Cor}[subsection]{Corollary}
\newtheorem{Lem}[subsection]{Lemma}
\newtheorem{Prop}[subsection]{Proposition}
\newtheorem{Conjec}[subsection]{Conjecture}

\newtheorem{Def}[subsection]{Definition}

\theoremstyle{remark}

\newtheorem{Rem}[subsection]{Remark}

\def\TeXref#1{%
        \leavevmode\vadjust{\setbox0=\hbox{{\tt
                \  {\tiny \textrm #1}}}%
        \theight=\ht0
        \advance\theight by \lineskip
        \kern -\theight \vbox to
        \theight{\rightline{\rlap{\box0}}%
        \vss}%
        }}%
\newif\ifShowLabels
\ShowLabelstrue
\newdimen\theight
\def\TeXrefEq#1{%
        \leavevmode\vadjust{\setbox0=\hbox{{\tt
                \  {\tiny \textrm #1}}}%
        \theight=\ht1
        \advance\theight by \lineskip
        \kern -\theight \vbox to
        \theight{\rightline{\rlap{\box0}}%
        \vss}%
        }}%

\newcommand{\refs}[1]{Section ~\ref{S:#1}}
\newcommand{\refss}[1]{Subsection ~\ref{SS:#1}}

\newcommand{\reft}[1]{Theorem ~\ref{T:#1}}
\newcommand{\refl}[1]{Lemma ~\ref{L:#1}}
\newcommand{\refp}[1]{Proposition ~\ref{P:#1}}
\newcommand{\refc}[1]{Corollary ~\ref{C:#1}}

\newcommand{\refd}[1]{Definition ~\ref{D:#1}}

\newcommand{\refe}[1]{\eqref{E:#1}}

\newenvironment{thm}[1]%
        { \begin{Thm} \label{T:#1}  \ifShowLabels \TeXref{T:#1} \fi }%
        { \end{Thm} }

\renewcommand{\th}[1]{\begin{thm}{#1}  }
\renewcommand{\eth}{\end{thm} }

\newenvironment{lemma}[1]%
        { \begin{Lem} \label{L:#1}  \ifShowLabels \TeXref{L:#1} \fi }%
        { \end{Lem} }

\newcommand{\lem}[1]{\begin{lemma}{#1} }
\newcommand{\elem}{\end{lemma}}

\newenvironment{propos}[1]%
        { \begin{Prop} \label{P:#1}  \ifShowLabels \TeXref{P:#1} \fi }%
        { \end{Prop} }

\newcommand{\prop}[1]{\begin{propos}{#1} }
\newcommand{\eprop}{\end{propos}}

\newenvironment{corol}[1]%
        { \begin{Cor} \label{C:#1}  \ifShowLabels \TeXref{C:#1} \fi }%
        { \end{Cor} }
\newcommand{\cor}[1]{\begin{corol}{#1}  }
\newcommand{\ecor}{\end{corol}}

\newenvironment{conjec}[1]%
        { \begin{Conjec} \label{Conj:#1}  \ifShowLabels \TeXref{C:#1} \fi }%
        { \end{Conjec} }
\newcommand{\conj}[1]{\begin{conjec}{#1}  }
\newcommand{\econj}{\end{conjec}}

\newenvironment{defeni}[1]%
        { \begin{Def} \label{D:#1}  \ifShowLabels \TeXref{D:#1} \fi }%
        { \end{Def} }
\newcommand{\defe}[1]{\begin{defeni}{#1}  }
\newcommand{\edefe}{\end{defeni}}

\newenvironment{remark}[1]%
        { \begin{Rem} \label{R:#1}  \ifShowLabels \TeXref{R:#1} \fi }%
        { \end{Rem} }
\newcommand{\rem}[1]{\begin{remark}{#1}}
\newcommand{\erem}{\end{remark}}

\newcommand{\eq}[1]%
        { \ifShowLabels \TeXrefEq{E:#1} \fi
           \begin{equation} \label{E:#1} }
\newcommand{\eeq}{\end{equation}}
\newcommand{\meq}[1]%
        { \ifShowLabels \TeXrefEq{E:#1} \fi
           \begin{multline} \label{E:#1} }
\newcommand{\emeq}{\end{multline}}

\newcommand{\prf}{ \begin{proof} }
\newcommand{\eprf}{ \end{proof} }
\newcommand{\Label}[1]{\label{#1}  \ifShowLabels \TeXref{#1} \fi }

\ShowLabelsfalse

\newcommand{\p}{\partial}

\newcommand{\Fp}{\operatorname{F.p._{s=0}}}

\newcommand{\LD}{\operatorname{log\,Det}}

\newcommand{\CLempty}{\operatorname{CL}}
\newcommand{\CL}{{\operatorname{CL}}(M,E)}
\newcommand{\CLm}{\CLempty_{(-1)}(M,E)}
\newcommand{\tCL}{\widetilde{\operatorname{CL}}(M,E)}
\newcommand{\tCLm}{\widetilde{\operatorname{CL}}_{(-1)}(M,E)}

\newcommand{\CLj}[1]{\operatorname{CL}^{#1}(M,E)}
\newcommand{\CLmj}[1]{\operatorname{CL}_{(-1)}^{#1}(M,E)}
\newcommand{\tCLj}[1]{\widetilde{\operatorname{CL}}^{#1}(M,E)}
\newcommand{\tCLmj}[1]{\widetilde{\operatorname{CL}}_{(-1)}^{#1}(M,E)}

\newcommand{\pdo}{pseudo-differential operator}

\newcommand{\ssub}[1]{_{(#1)}}
\newcommand{\stet}{_{(\tet)}}

\newcommand{\stetmpi}{_{(\tet-m\pi)}}
\newcommand{\tilPi}{\widetilde{\Pi}}

\newcommand{\TR}{\operatorname{TR}}
\newcommand{\sym}{{\operatorname{sym}}}

\begin{document}



\title[Symmetrized Trace and Symmetrized Determinant]{Symmetrized Trace and Symmetrized Determinant of Odd Class Pseudo-Differential Operators}
\ \author{Maxim Braverman}
\address{Department of Mathematics\\
        Northeastern University   \\
        Boston, MA 02115 \\
        USA
         }
\email{maximbraverman@neu.edu}

\begin{abstract}
We introduce a new canonical trace on odd class logarithmic pseudo-differential operators on an odd dimensional manifold, which vanishes on
commutators. When restricted to the algebra of odd class classical pseudo-differential operators our trace coincides with the canonical trace of
Kontsevich and Vishik. Using the new trace we construct a new determinant of odd class classical elliptic pseudo-differential operators. This
determinant is multiplicative up to sign whenever the multiplicative anomaly formula for usual determinants of Kontsevich-Vishik and Okikiolu holds.
When restricted to operators of Dirac type our determinant provides a sign refined version of the determinant constructed by Kontsevich and Vishik.
We discuss some applications of the symmetrized determinant to a non-linear $\sigma$-model in superconductivity.
\end{abstract}

\maketitle
\section{Introduction}\Label{S:introd}

To define a trace and a determinant of an unbounded (pseudo-)differential operator one needs to use a regularization procedure, which usually
leads to so called anomalies in the behavior of the obtained regularized trace and determinant. Namely
\begin{itemize}
\item The obtained trace and determinant depend on the choices made for the regularization.
\item The regularized trace is not tracial, i.e., does not vanish on commutators.
\item The regularized determinant is not multiplicative, i.e., $\Det{}AB\not= \Det A\cdot\Det B$.
\end{itemize}
It is well known that, in general, those anomalies are unavoidable. For example, there is no extension of the usual trace of trace class operators to the algebra of
all pseudo-differential operators, which is tracial.  However, if one restricts to a subalgebra one can hope to have a tracial extension of the usual trace, cf.
\cite{Ducourtioux99}. In particular, Kontsevich and Vishik \cite[\S4]{KontsevichVishik_long} constructed a tracial functional $\Tr_{(-1)}$ on the algebra of odd class
classical \pdo s on an odd dimensional manifold.

The usual regularization procedure depends on the choice of a spectral cut of the complex plane. In this paper we propose a new regularized
trace and determinant, called the {\em symmetrized trace} and the {\em symmetrized determinant}, of odd class operators on an odd dimensional
manifold, obtained by averaging the usual definitions for different spectral cuts. The symmetrized trace is independent of any choices. The
symmetrized determinant still depends on the spectral cut, but this dependence is in a sense weaker than for the usual regularized trace and
determinant. The symmetrized trace is tracial and the symmetrized determinant is very often multiplicative. We will now explain our
constructions in some more details.

\subsection{The symmetrized trace}\Label{SS:Isymtrace}
Fix an elliptic pseudo-differential operator $Q$ of a positive order $m$ and assume that $\tet$ is an Agmon angle for $Q$, cf. \refss{complexpower}. The {\em
$Q$-weighted trace} $\Tr^Q_{(\tet)}A$ of a logarithmic pseudo-differential operator $A$ is defined to be the constant term in the Laurent expansion at $s=0$ of the
function $s\mapsto \TR(AQ_{(\tet)}^s)$. Here, $Q_{(\tet)}^s$ is the $s$-th power of $Q$ defined using the spectral cut $\tet$ and $\TR$ is the Kontsevich-Vishik
canonical trace, introduced in \cite[\S3]{KontsevichVishik_short}, \cite[\S3]{KontsevichVishik_long} for classical \pdo s, and generalized in \cite[\S5]{Lesch99}. The
weighted trace depends on both, $Q$ and $\tet$, and, in general, is not tracial.

Suppose now that the dimension of the manifold is odd, that both operators $Q$ and $A$ are of odd class, cf. \refs{preliminaries}, and that both, $\tet$ and
$\tet-m\pi$, are Agmon angles for $Q$. Then we define the {\em symmetrized trace} $\Tr^{\sym}A$ of $A$ by the formula
\[
    \Tr^{\sym}A \ := \ \frac12\,\Big(\,\Tr_{(\tet)}^QA\,+\,\Tr^Q_{(\tet-m\pi)}A\,\Big).
\]
Then, cf. \refp{Trsym}, the symmetrized trace is independent of the choice of $Q$ and $\tet$. If $A$ is a classical \pdo, then $\Tr^{\sym}A$
coincides with the Kontsevich-Vishik canonical trace $\Tr_{(-1)}A$, cf. \refss{canonicaltrace}. Recall that Kontsevich and Vishik defined
$\Tr_{(-1)}A$  using an even order positive definite operator $Q$. Thus we also obtain a new formula for $\Tr_{(-1)}A$, which defines it using
an operator which is not necessarily positive definite.

To define a determinant of $A$, one needs to define a trace of the logarithm of $A$, which is not a classical \pdo. Hence, we are mostly
interested in the properties of the symmetrized trace on the space of logarithmic \pdo s. Our main result here is \reft{tracial}, which claims
that $\Tr^{\sym}$ is tracial in the sense that
\[
  \Tr^{\sym}[A,B]\ = \ 0,
\]
for any odd class logarithmic \pdo s $A$ and $B$.

\subsection{The symmetrized determinant}\Label{SS:Isymdet}
Assume now that $A$ is an odd class elliptic classical \pdo\ of positive integer order $m$ and that both, $\tet$ and $\tet-m\pi$, are Agmon angles for $A$. We define
the {\em symmetrized determinant} $\Det^\sym_{(\tet)}A$ of $A$ by the formula
\eq{Isymdet}
        \LD^\sym\stet A \ := \ \frac12\,\Tr^{\sym}\,\big(\log\stet A+\log\stetmpi A\big).
\end{equation}
If the order $m$ of $A$ is even, then the symmetrized determinant \refe{Isymdet} is equal to the usual  $\zet$-regularized determinant of $A$, cf.
\refp{detsymeven}, but in the case when $m$ is odd, the symmetrized determinant might be quite different from the usual one. By \refp{detsymodd}, in
this case
\eq{IdetA2}
     \big(\Det_{(\tet)}^\sym{}A\big)^2\ = \ \Det^{\sym}_{(2\tet)}(A^2).
\end{equation}
Combining Propositions~\ref{P:detsymeven} and \ref{P:detsymodd} we see that up to sign the symmetrized determinant is quite a classical object.
Definition \refe{Isymdet} fixes a sign which is crucial for applications in Subsections~\ref{SS:Irealcoef} and \ref{SS:Isigma}. Also our proof
of the multiplicativity of the symmetrized determinant relies heavily on the construction of the symmetrized trace and formula \refe{Isymdet}.

As the usual $\zet$-regularized determinant, the symmetrized determinant does depend on the spectral cut $\tet$, but this dependence is weaker
than in the case of the usual determinant, cf. \refp{deponangle}. In particular, if the order of $A$ is odd and its leading symbol is
self-adjoint, then $\Det^\sym_{(\tet)}A$ is independent of $\tet$ up to sign.

Our main result is \reft{multprop}. Roughly speaking it says that whenever the multiplicative anomaly formula of Kontsevich-Vishik and Okikiolu for
usual determinants  holds, the symmetrized determinant is multiplicative up to sign. For example, cf. \refc{multcom}, if the leading symbols of the
operators $A$ and $B$ commute and the leading symbol of $A$ is self-adjoint, then
\[
    \Det^{\sym}_{(\tet_B+\eps \tet_A)}AB \ = \ \pm\,\Det^\sym_{(\tet_A)}A\cdot \Det^\sym_{(\tet_B)}B,
\]
where $\tet_A,\tet_B\in (0,2\pi)$ are Agmon angles for the operators $A$ and $B$ respectively and $\eps=1$ if $\tet_A\in (0,\pi)$ and $\eps=-1$ if
$\tet_A\in (\pi,2\pi)$.

In the case when the operator $A$ is of Dirac type Kontsevich and Vishik, \cite[\S4.1]{KontsevichVishik_long}, suggested to a define  a new
determinant of $A$ as a square root of the determinant of the Laplace-type operator $A^2$. Thus their definition had a sign indeterminacy. It
follows from the equation \refe{IdetA2} that the reduction modulo signs of the symmetrized determinant $\Det\stet^\sym{}A$ coincides in the case
of Dirac-type operators with the Kontsevich-Vishik determinant. Thus in the case of Dirac-type operators our symmetrized determinant provides a
more direct construction of the Kontsevich-Vishik determinant and also fixes the signs in its definition.

\subsection{Operators with spectrum symmetric about the real axis}\label{SS:Irealcoef}
Let $A$ be an odd class classical \pdo\ of odd order. Assume that the leading symbol of $A$ is self-adjoint with respect to a Hermitian metric on $A$. Suppose further
that the spectrum of $A$ is symmetric about the real axis, cf. \refd{realcoef}. Note that any differential operator with real coefficients has this property. We
denote by $m_+$ the number of the eigenvalues of A (counting with their algebraic multiplicities) which lie on the positive part of the imaginary axis (since the
spectrum of $A$ is symmetric $m_+$ is equal to the number $m_-$ of the eigenvalues which lie on the negative part of the imaginary axis). In \reft{realcoef} we show
that {\em the symmetrized determinant $\Det^\sym\stet{}A$ is real and its sign is equal to $(-1)^{m_+}$, i.e.,}
\eq{Irealcoef}
    \Det^\sym\stet A \ = \ (-1)^{m_+}\,\big|\,\Det^\sym\stet A\,\big|.
\end{equation}
Note that this result is somewhat surprising, since for a finite dimensional matrix whose spectrum is symmetric about the real axis the sign of the determinant is
independent of the number of imaginary eigenvalues and is determined by the number of eigenvalues which lie on the negative part of the real axis.

The equation \refe{Irealcoef} should be compared with the main theorem of \cite{BrAbanov} where a similar equality was obtained for the usual $\zet$-regularized
determinant of an operator whose spectrum is symmetric about the imaginary axis.

\subsection{Application to a non-linear $\sig$-model}\label{SS:Isigma}
In the recent years several examples appeared in physical literature when the phase of the determinant of a geometrically defined non self-adjoint Dirac-type operator
is a topological invariant, see e.g., \cite{Redlich84,Redlich84a,AbWie-chir,AbWie-geom,AbanovWieg00,Abanov-Hopf}. In some of these examples the physicists claim that
the determinant is real and compute its sign. Unfortunately, their arguments are not rigorous. In particular, they never specify which spectral cut they use to define
the determinant. In \cite{BrAbanov} we tried to better understand this phenomenon. In particular, we provided a rigorous computation of the sign of roughly one half
of the examples from \cite{AbanovWieg00,Abanov-Hopf} in which the spectrum of the Dirac-type operator is symmetric about the imaginary axis. Unfortunately, the
methods of \cite{BrAbanov} are not applicable to the other half of the examples in which the spectrum is symmetric about the real axis, cf. the operator
$\Gam\circ{}D_{mn}$ in section~6.3 of \cite{BrAbanov}. Moreover, one can show that the usual $\zet$-regularized determinant is not real in some of these examples and
the computations of physicists are not legal, essentially because they don't take into account the effect of the choice of the spectral cut.\footnote{More precisely,
they consider a deformation $A(t)$ of $A$ and don't take into account the fact that one can not take the same spectral cut for all values of $t$.} There is, however,
no reason to believe that the usual $\zet$-function regularization of the determinant is the most adequate for the description of physics. Moreover, it turns out that
the $\sig$-model constructed using the usual $\zet$-regularized determinant fails to satisfy some basic conservation laws.  One of our main motivation for the study
of the symmetrized determinants was an attempt to give a rigorous description of the results of \cite{AbanovWieg00,Abanov-Hopf}. We will discuss the obtained
$\sig$-models elsewhere. In particular, we will explain that \refe{Irealcoef} gives the  sign of the determinant predicted in \cite{AbanovWieg00,Abanov-Hopf}.

\section{Preliminaries}\label{S:preliminaries}

\subsection{Classical pseudo-differential operators of odd class}\label{SS:classodd}
Let $M$ be a closed $d$-dimensional manifold and let $E$ be a complex vector bundle over $M$. We denote by $\CLj{m}$ the space of order $m\in
\CC$ classical pseudo-differential operators
\[
    A:\,C^\infty(M,E)\ \longrightarrow\ C^\infty(M,E),
\]
cf. \cite{ShubinPDObook}. Recall that for each $A\in \CLj{m}$ the
symbol $\sig(A)(x,\xi)$ has an asymptotic expansion of the form
\eq{CL}
    \sig(A)(x,\xi) \ \sim \ \sum_{j=0}^\infty\, \sig_{m-j}(A)(x,\xi), \qquad\qquad (x,\xi)\in T^*M,
\end{equation}
where each $\sig_{m-j}(x,\xi)$ is {\em positive homogeneous} in $\xi$ of degree $m-j$, i.e.
\eq{homm-j}
    \sig_{m-j}(A)(x,t\xi) \ = \ t^{m-j}\,\sig_{m-j}(A)(x,\xi), \qquad \text{for every}\quad t>0.
\end{equation}
The functions $\sig_{m-j}(A)(x,\xi)$ are called the {\em positive homogeneous components} of the symbol of $A$. The function $\sig_m(x,\xi)$ is
called the {\em principal symbol} of $A$.

Set $\CL=\bigcup_{m\in \CC}\CLj{m}$.

An {\em odd class} operator of an {\em integer} order $m$ is an operator $A\in \CLj{m}$ such that
\eq{oddcondition}
    \sig_{m-j}(A)(x,-\xi) \ = \ (-1)^{m-j}\,\sig_{m-j}(A)(x,\xi), \qquad\qquad j=0,1\ldots
\end{equation}
We denote by $\CLmj{m}$ the space of odd class operators of order $m\in \ZZ$ and we set
\eq{CLm}
    \CLm\ =\ \bigcup_{m\in\ZZ}\,\CLmj{m}.
\end{equation}
Note that all {\em differential} operators belong to $\CLm$.

\subsection{Logarithmic differential operators}\label{SS:logaritmoper}
We say that a \pdo\ $A:C^\infty_c(M,E)\to C^\infty(M,E)$ is {\em logarithmic} of degree $m\in \CC$, cf. \cite{Okikiolu95CH,Okikiolu95MA}, if its
symbol has an asymptotic expansion of the form
\eq{tCL}
    \sig(A)(x,\xi) \ \sim \ \gam\,\log|\xi| \ + \ \sum_{j=0}^\infty\, \sig_{m-j}(A)(x,\xi), \qquad\qquad (x,\xi)\in T^*M,\ m,\gam\in \CC,
\end{equation}
where each $\sig_{m-j}(x,\xi)$ is positive homogeneous in $\xi$ of degree $m-j$. The number $m\in \CC$ is called the degree of the logarithmic
\pdo\ $A$, and the number $\gam\in \CC$ is called the {\em type} of this operator. We denote the set of logarithmic \pdo s of degree $m$ and
type $\gam$  ($(m,\gam)$-logarithmic, for short) by $\tCLj{m,\gam}$ and we set
\eq{tCL2}
    \tCLj{m} \ = \ \bigcup_{\gam\in\CC}\,\tCLj{m,\gam}, \qquad \tCL\ =\ \bigcup_{m\in\ZZ}\,\tCLj{m}.
\end{equation}
Note that logarithmic pseudo-differential operators are not classical.

We say that a logarithmic \pdo\ $A\in \tCLj{m}$ $(m\in\ZZ)$ is of {\em odd class} if in the asymptotic expansion \refe{tCL} each term $\sig_{m-j}(A)(x,\xi)$ satisfies
\refe{oddcondition}. Denote by $\tCLm$ (resp. $\tCLmj{m,\gam}$) the set of odd class logarithmic (resp. odd class $(m,\gam)$-logarithmic) \pdo s.

Using the standard rules of composition of symbols (cf., for example, \cite[Th.~3.5]{ShubinPDObook}, \cite[\S1]{Okikiolu95MA}) we immediately
get
\lem{[logA,B]}
(i)\ If\/ $A, B\in \tCL$,  then the commutator $[A,B]= AB-BA$ is a classical \pdo, $[A,B]\in \CL$.

(ii)\ If\/ $A, B\in \tCLm$, then $[A,B]\in \CLm$.
\elem

\subsection{Odd pair of logarithmic \pdo s}\label{SS:oddpair}
We say that the operators $A,B\in \tCLj{m}$ form an {\em odd pair} (or that the pair $(A,B)$ is odd) if, for all $j=0,1,\ldots$,
\eq{oddpair}
    \sig_{m-j}(A)(x,-\xi)\ = \ (-1)^{m-j}\,\sig_{m-j}(B)(x,\xi),
\end{equation}
where $\sig_{m-j}(A)(x,\xi)$, $\sig_{m-j}(B)(x,\xi)$ are positive homogeneous components of the symbols of $A$ and $B$ respectively, cf.
\refe{tCL}.

Clearly, if $A\in \tCLm$ then the pair $(A,A)$ is odd. Further, if $(A,B)$ is an odd pair, then $A+B\in \tCLm$.

From the standard formulae of composition of symbols (cf., for example, \cite[Th.~3.5]{ShubinPDObook}, \cite[\S1]{Okikiolu95MA}) we immediately
obtain
\lem{oddtimesodd}
Given two odd pairs $(A_1,B_1)$ and $(A_2,B_2)$ of\/ {\em classical} \pdo s, the pair \/ $(A_1A_2,B_1B_2)$ is also odd.
\elem

\subsection{Complex powers of elliptic \pdo s}\label{SS:complexpower}
An angle $\tet\in [0,2\pi)$ is said to be a {\em principal angle} for a \pdo\ $A\in \CLj{m}$ ($m\in \ZZ_+$) if there exists a conical
neighborhood $\Lam$ of the ray
\eq{Rtet}
    R_\tet\ := \ \big\{\,\rho\,e^{i\tet}:\ \rho\ge0\,\big\}
\end{equation}
such that, for each $\xi\not=0$, the principal symbol $\sig_m(A)(x,\xi)$ of $A$ does not have eigenvalues in $\Lam$. In particular, the
existence of a principal angle implies that the operator $A$ is elliptic and that the spectrum of $A$ is discrete.

An {\em Agmon angle} for $A$ is a principal angle such that there are no eigenvalues of $A$ on the ray $R_\tet$. If $A$ admits an Agmon angle
then it is elliptic and invertible. In this situation there exists $\eps>0$ such that there are no eigenvalues of $A$ in the solid angle
\[
    \Lam_{[\tet-\eps,\tet+\eps]} \ := \ \big\{ \rho\,e^{i\alp}:\ \rho\ge0,\, \tet-\eps\le\alp\le\tet+\eps\,\big\}.
\]
Suppose $A\in \CLj{m}$ ($m\in \ZZ_+$) admits an Agmon angle $\tet$. Then, cf. \cite{Seeley67}, \cite[Ch.~II,\S8]{ShubinPDObook}, for each $s\in
\CC$ a complex power $A\stet^s\in \CLj{sm}$ is defined. The symbol of $A\stet^s$ can be calculated as follows: Let $R(\lam)= (A-\lam)^{-1}$
(where $\lam\in \Lam_{[\tet-\eps,\tet+\eps]}$) denote the resolvent of $A$. The symbol $r(x,\xi;\lam)$ of $R(\lam)$ has an asymptotic expansion
of the form
\eq{rxxi}
    r(x,\xi;\lam) \ \sim \ \sum_{j=0}^\infty\, r_{-m-j}(x,\xi;\lam),
\end{equation}
where the terms $r_{k}(x,\xi;\lam)$ are positive homogeneous in the sense that, for $t>0$
\eq{rhomog}
    r_k\big(x,t\xi;t^{m}\lam\big) \ = \ t^k\,r_k(x,\xi;\lam), \qquad (x,\xi)\in T^*M,\ \lam\in \Lam_{[\tet-\eps,\tet+\eps]}.
\end{equation}
Moreover, if $A\in \CLmj{m}$, then it follows from the explicit formulae for $r_k(x,\xi;\lam)$, cf. \cite[\S11.1]{ShubinPDObook},
\cite[\S2]{KontsevichVishik_long}, that
\eq{rodd}
    r_k\big(x,-\xi;(-1)^m\lam\big) \ = \ (-1)^k\,r_k(x,\xi;\lam), \qquad (x,\xi)\in T^*M,\ \lam\in \Lam_{[\tet-\eps,\tet+\eps]}.
\end{equation}

The symbol $\sig(A\stet^s)(x,\xi)$ has the asymptotic expansion, cf. \cite[\S11.2]{ShubinPDObook}, \cite[\S2]{KontsevichVishik_long},
\eq{sigAs}
    \sig(A\stet^s)(x,\xi) \ \sim \ \sum_{j=0}^\infty\, \sig_{sm-j}(A\stet^s)(x,\xi),
\end{equation}
where the terms $\sig_{sm-j}(A\stet^s)(x,\xi)$ are positive homogeneous of degree $sm-j$. They depend analytically on $s\in\CC$ and for
$\RE{}s<0$ are given by
\eq{sigAsj}
    \sig_{sm-j}(A\stet^s)(x,\xi) \ = \ \frac{i}{2\pi}\, \int_{\Gam\stet}\, \lam\stet^s\,r_{-m-j}(x,\xi;\lam)\,d\lam,
\end{equation}
where the contour $\Gam\stet= \Gam_{(\tet),\rho_0}\subset \CC$ consisting of three curves $\Gam= \Gam_1\cup \Gam_2\cup \Gam_3$,
\begin{gather}\Label{E:Gamtetrho}\notag
    \Gam_1 \ = \ \big\{\, \rho e^{i\tet}:\, \infty >\rho\ge \rho_0\, \big\},
    \quad
    \Gam_2 \ = \ \big\{\, \rho_0 e^{i\alp}:\, \tet> \alp>\tet-2\pi\, \big\},\\
    \quad
    \Gam_3 \ = \ \big\{\, \rho e^{i(\tet-2\pi)}:\, \rho_0\le \rho<\infty\, \big\}.
\end{gather}
Here $\rho_0>0$ is a small enough number, and $\lam\stet^s$ is defined as $e^{s\log\stet\lam}$, where $\tet\ge\IM\log\stet\lam\ge \tet-2\pi$ (i.e., $\IM\log\stet\lam=
\tet$ on $\Gam_1$ and $\IM\log\stet\lam= \tet-2\pi$ on $\Gam_3$).

\lem{Asodd}
Suppose $A\in \CLmj{m}$ and that both, $\tet$ and $\tet-m\pi$, are Agmon angles for $A$. Then, for each $j=0,1,\ldots$, $(x,\xi)\in T^*M$, $s\in
\CC$,
\eq{Asoddodd}
    \sig_{sm-j}(A\stet^s)(x,-\xi) \ = \ (-1)^j\,e^{ims\pi}\, \sig_{sm-j}(A\stetmpi^s)(x, \xi).
\end{equation}
\elem
\prf
Assume, first, that $\RE{}s<0$. Then, using \refe{rodd}, we obtain
\meq{Asoddodd2}\notag
    \sig_{sm-j}(A\stet^s)(x,-\xi) \ = \ \frac{i}{2\pi}\, \int_{\Gam\stet}\, \lam\stet^s\,r_{-m-j}(x,-\xi;\lam)\,d\lam\\
        \ = \ (-1)^{m+j}\,\frac{i}{2\pi}\, \int_{\Gam\stet}\, \lam\stet^s\,r_{-m-j}(x,\xi;(-1)^m\lam)\,d\lam\\
        \ = \ (-1)^{m+j}\,\frac{i}{2\pi}\, \int_{\Gam\stetmpi}\, (e^{im\pi}\mu)\stet^s\,r_{-m-j}(x,\xi;\mu)\,d(e^{im\pi}\mu) \\
        \ = \ (-1)^{m+j}\,e^{im\pi}\,\frac{i}{2\pi}\, \int_{\Gam\stetmpi}\, e^{ims\pi}\,\mu\stetmpi^s\,r_{-m-j}(x,\xi;\mu)\,d\mu
        \\ = \ (-1)^j\,e^{ims\pi}\, \sig_{sm-j}(A\stetmpi^s)(x, \xi).
\end{multline}
Since both, the left and the right hand side of this equality, are analytic in $s$, we conclude that the equality holds for all $s\in \CC$.
\eprf

\subsection{Logarithms of elliptic \pdo s}\label{SS:logarithm}
For $A\in \CLj{m}$ admitting an Agmon angle $\tet$, the logarithm $\log\stet{A}$ of $A$ is defined by the formula
\eq{logA}
    \log\stet A \ = \ \frac{\p}{\p s}\Big|_{s=0}A\stet^s.
\end{equation}
Then, cf. \cite[\S2]{KontsevichVishik_long}, \cite[Lemma~2.4]{Okikiolu95CH}, $\log\stet{}A\in \tCLj{0,m}$. In particular, the symbol of
$\log\stet{}A$ admits an asymptotic expansion
\eq{symlogA}
    \sig(\log\stet{}A)(x,\xi) \ \sim \ m\,\log|\xi| \ + \ \sum_{j=0}^\infty\, \sig_{-j}(\log\stet{}A)(x,\xi),
\end{equation}
and its positive homogeneous components $\sig_{-j}(\log\stet{}A)(x,\xi)$ are given by the formulae
\eq{symlogA=}
    \sig_{-j}(\log\stet{}A)(x,\xi) \ = \ |\xi|^{-j}\,\p_s\,\sig_{sm-j}(A\stet^s)(x,\xi/|\xi|)|_{s=0}, \qquad \text{for}\quad j\ge0.
\end{equation}

From \refe{tCL} and \refe{symlogA} we immediately obtain the following
\lem{log=log+A}
Let $B\in \CLj{m}$ be a classical elliptic \pdo\ of positive order and suppose that $\tet$ is an Agmon angle for $B$. Then for any logarithmic
\pdo\ $A\in \tCLj{n,m}$ the difference $A-\log\stet{}B\in \CLj{n}$. In particular, any $A\in \tCL$ can be represented in the form
\eq{log=log+A}
    A\ =\ A_1\ +\ \log\stet{}A_2,
\end{equation}
where $A_1,\ A_2$ are classical \pdo s.
\elem

\subsection{Dependence of logarithm on the angle}\label{SS:angle}
Let $\Pi_{(\tet_1,\tet_2)}$ denote the spectral projection of $A$ corresponding to the eigenvalues which lie in the solid angle
\[
    \Lam_{(\tet_1,\tet_2)} \ := \ \big\{\,\rho\,e^{i\alp}:\, \rho\ge0, \ \tet_1<\alp<\tet_2\,\big\}.
\]
In particular, $\Pi_{(\tet_1,\tet_2)}= \Id$ if $|\tet_2-\tet_1|\ge 2\pi$.

Suppose now that $\tet_1, \tet_2$ are Agmon angles for $A$. It follows immediately from the definition \refe{logA} of the logarithm that if $2k\pi\le\tet_2-\tet_1<
2(k+1)\pi$ for some integer $k$ then, cf.  formula (1.4) of \cite{Okikiolu95CH},
\eq{loglogtet1tet2}
    \log_{(\tet_2)} A \ = \ 2ik\pi\,\Id \ + \ 2i\pi\,\Pi_{\tet_1,\tet_2-2k\pi} \ + \ \log_{(\tet_1)}A.
\end{equation}
In particular, if both, $\tet$ and $\tet-2k\pi$  ($k\in \ZZ$) are Agmon angles for $A$, then
\eq{loglogpair2kpi}
    \log\stet A \ = \ 2ik\pi\,\Id  \ + \ \log_{(\tet-2k\pi)} A.
\end{equation}

If $A\in \CL$ is a classical \pdo\ we denote by $\Res{}A$ the non-commutative residue of Wodzicki \cite{Wodzicki84,Wodzicki87} and Guillemin
\cite{Guillemin85} (see also \cite{Kassel99} for a review).
\lem{reslog}
Suppose $A_j\in \CLj{m_j}$ ($j=1,2$) are classical elliptic \pdo s of positive orders. Then the number
\[
    \Res\Big(\, \frac{\log\ssub{\tet_1}A_1}{m_1}-\frac{\log\ssub{\tet_2}A_2}{m_2}\,\Big)
\]
is independent of the choice of the Agmon angles $\tet_1$ and $\tet_2$.
\elem
Note that the operator $\frac{\log\ssub{\tet_1}A_1}{m_1}-\frac{\log\ssub{\tet_2}A_2}{m_2}$ is classical and hence its residue is well defined.
\prf
Let $\tet_j'\in [2k_j\pi+\tet_j,2(k_j+1)\pi+\tet_j]$  be another Agmon angle for $A_j$ \ ($j=1,2$). Let $\Pi_{\tet_j,\tet_j'-2k_j\pi}(A_j)$
denote the spectral projection of $A_j$ corresponding to the solid angle $\Lam_{\tet_j,\tet_j'-2k_j\pi}$. Then it follows from
\refe{loglogtet1tet2} that
\meq{loglog-loglog}
 \Big(\, \frac{\log\ssub{\tet_1}A_1}{m_1}-\frac{\log\ssub{\tet_2}A_2}{m_2}\,\Big) \ - \
 \Big(\, \frac{\log\ssub{\tet_1'}A_1}{m_1}-\frac{\log\ssub{\tet_2'}A_2}{m_2}\,\Big) \\ = \
 2i\pi\big(\,\frac{k_1}{m_1}-\frac{k_2}{m_2}\,\big)\,\Id \ + \ \frac{2i\pi}{m_1}\,\Pi_{\tet_1,\tet_1'-2k_1\pi}(A_1) \ - \
    \frac{2i\pi}{m_2}\,\Pi_{\tet_2,\tet_2'-2k_2\pi}(A_2).
\end{multline}
Since the Wodzicki residue of a pseudo-differential projection vanishes, cf. \cite[\S6]{Wodzicki84}, \cite{Wodzicki87}, the residue of the right hand side of
\refe{loglog-loglog} is equal to 0.
\eprf

\subsection{Logarithms of odd class elliptic \pdo s}\label{SS:logarithmodd}
From equality \refe{symlogA=} and \refl{Asodd} we obtain the following
\lem{logAoddeven}
Suppose $A\in \CLmj{m}$ and both, $\tet$ and $\tet-m\pi$, are Agmon angles for $A$, then
\eq{loglogpair}
     \big(\,\log\stet{A},\, im\pi\Id+\log\stetmpi{A}\,\big)
\end{equation}
is an odd pair of operators, cf. \refss{oddpair}.

In particular, if\/ $m$ is even, then $\log\stet{}A\in \tCLmj{0,m}$.
\elem
\prf
Since $A\stet^s\big|_{s=0}$ is the identity operator, the positive homogeneous components of its symbol are given by
\eq{A0}
    \sig_{sm-j}(A\stet^s)(x,\xi)|_{s=0} \ = \ \del_{j,0}\,\Id.
\end{equation}
Thus, from \refe{Asoddodd} we obtain
\eq{logAj}
    \p_s\,\sig_{sm-j}(A\stet^s)(x,-\xi/|\xi|)|_{s=0} \ = \ im\pi\,\del_{j,0} \ + \ (-1)^j\,\p_s\,\sig_{sm-j}(A\stetmpi^s)(x,\xi/|\xi|)|_{s=0}.
\end{equation}
Hence, by \refe{symlogA=}, the pair \refe{loglogpair} is odd.

If $m$ is even, then from \refe{loglogpair2kpi} and \refe{loglogpair} we conclude now that the pair $(\log\stet{}A,\log\stet{}A)$ is odd, which,
by definition of the odd pair, means that $\log\stet{}A\in \tCLmj{0,m}$.
\eprf
\cor{logAoddodd}
Let\/ $A\in \CLmj{m}$ and both, $\tet$ and $\tet-m\pi$, are Agmon angles for $A$. Then
\[
    \log\stet A \ + \ \log\stetmpi A \ \in \ \tCLmj{0,m}.
\]
\ecor

\prop{reslog-log}
Suppose $A_j\in \CLmj{m_j}$ ($j=1,2$) are odd class classical elliptic \pdo s of positive orders on an odd dimensional manifold $M$. Then
\eq{reslog-log}
    \Res\Big(\, \frac{\log\ssub{\tet_1}A_1}{m_1}-\frac{\log\ssub{\tet_2}A_2}{m_2}\,\Big) \ = \ 0.
\end{equation}
for any Agmon angles $\tet_1$ and $\tet_2$.
\eprop
\prf
By \refl{reslog} the left hand side of \refe{reslog-log} is independent of $\tet_1,\tet_2$. Hence, we can assume that $\tet_1,\tet_2$ are chosen
so that the angles $\tet_1-m_1\pi, \tet_2-m_2\pi$ are also Agmon angles for operators $A_1, A_2$ respectively. From \refc{logAoddodd} we see
that
\[
  \Big(\, \frac{\log\ssub{\tet_1}A_1}{m_1}-\frac{\log\ssub{\tet_2}A_2}{m_2}\,\Big) \ + \
  \Big(\, \frac{\log\ssub{\tet_1-m_1\pi}A_1}{m_1}-\frac{\log\ssub{\tet_2-m_2\pi}A_2}{m_2}\,\Big) \ \in \ \CLmj{0}.
\]
Since the Wodzicki residue of an odd class operator on an odd dimensional manifold vanishes, cf. \cite[Lemma~7.3]{KontsevichVishik_short},
\cite[Remark~4.5]{KontsevichVishik_long}, it follows that
\eq{Reslog-log=-Reslog-log}
    \Res\Big(\, \frac{\log\ssub{\tet_1}A_1}{m_1}-\frac{\log\ssub{\tet_2}A_2}{m_2}\,\Big) \ = \
   -\Res\Big(\, \frac{\log\ssub{\tet_1-m_1\pi}A_1}{m_1}-\frac{\log\ssub{\tet_2-m_2\pi}A_2}{m_2}\,\Big).
\end{equation}
On the other side, by \refl{reslog}
\eq{Reslog-log=Reslog-log}
    \Res\Big(\, \frac{\log\ssub{\tet_1}A_1}{m_1}-\frac{\log\ssub{\tet_2}A_2}{m_2}\,\Big) \ = \
   \Res\Big(\, \frac{\log\ssub{\tet_1-m_1\pi}A_1}{m_1}-\frac{\log\ssub{\tet_2-m_2\pi}A_2}{m_2}\,\Big).
\end{equation}
The equality \refe{reslog-log} follows immediately form \refe{Reslog-log=-Reslog-log} and \refe{Reslog-log=Reslog-log}.
\eprf

\section{The symmetrized trace}\label{D:symtrace}

In the first part of this section we recall the notion of {\em weighted trace} of a \pdo\ and discuss its basic properties, cf.
\cite{MelroseNistor96,CardonaDucourtiouxMagnotPaycha02,Ducourtioux99}. Then we define a new canonical trace, called the  {\em symmetrized trace}
$\Tr^{\sym}$, of a logarithmic \pdo\ of {\em odd class} on an {\em odd dimensional manifold} and show that it is {\em tracial}, in the sense
that $\Tr^{\sym}[A,B]= 0$ for all $A, B\in \tCLm$. For odd class {\em classical} \pdo s the symmetrized trace coincides with the
Kontsevich-Vishik canonical trace $\Tr_{(-1)}$ introduced in \cite{KontsevichVishik_long,KontsevichVishik_short} by a slightly different
procedure. Thus our symmetrized trace extends the Kontsevich-Vishik canonical trace to logarithmic \pdo s of odd class.

\subsection{Weighted traces}\label{waightedtr}
Let $Q\in \CL$ be a classical elliptic \pdo\ of positive order admitting an Agmon angle $\tet$. For $A\in \tCL$, consider the generalized
$\zet$-function
\eq{zetQ}
    s \ \mapsto \ \zet_\tet(s,Q;A) \ := \ \TR\,(AQ\stet^s),
\end{equation}
where $\TR$ stands for the Kontsevich-Vishik canonical trace, introduced in \cite[\S3]{KontsevichVishik_short},
\cite[\S3]{KontsevichVishik_long} for classical \pdo s, and generalized in \cite[\S5]{Lesch99}. For $\RE s\ll-1$, the operator $AQ\stet^s$ is of
trace class and the Kontsevich-Vishik trace coincides with the usual trace.
\lem{meromorpic}
The function \refe{zetQ} is meromorphic in $s$ and has at most simple pole at $s=0$.
\elem
\prf
For $A\in \CL$ it is shown in \cite[Th.~3.1]{KontsevichVishik_short}, \cite[Prop.~3.4]{KontsevichVishik_long}. For $A= \log\stet{}B$ the
statement of the lemma is proven in Prop.~2 of \cite{Ducourtioux99}. The general case, follows from this two special cases and the existence of
the decomposition \refe{log=log+A}.
\eprf

We shall use the following notations: Suppose $f(s)$ is a function of a complex parameter $s$ which is meromorphic near $s=0$. We call the zero
order term in the Laurent expansion of $f$ near $s=0$ the {\em finite part of $f$ at 0} and denote it by $\Fp f(s)$.

\defe{weightedtr}
Let $Q\in \CL$ be a classical elliptic  \pdo\ of positive order admitting an Agmon angle $\tet$. The {\em $Q$-weighted trace} $\Tr^Q{}A$ of a
logarithmic \pdo\ $A\in \tCL$ is defined as
\eq{weightedtr}
    \Tr^Q\stet{}A \ := \  \Fp\,\TR\,(AQ\stet^s).
\end{equation}
\edefe
A slightly more explicit expression for $\Tr^Q\stet{}A$ is obtained in Definitions~3 and 4  of \cite{Ducourtioux99}.

Note that though traditionally we call \refe{weightedtr} a trace it is not a trace on the algebra of \pdo s since it does not vanishes on
commutators. More precisely, Proposition~1 of \cite{CardonaDucourtiouxMagnotPaycha02} states that
\prop{notatracre}
Suppose $Q\in \CLj{m}$ is a classical elliptic \pdo\ of positive order admitting an Agmon angle $\tet$. Then for any classical \pdo\ $A,B\in
\CL$
\eq{notatrace}
    \Tr^Q\stet{}[A,B] \ = \ -\frac1m\,\Res\big(\,[\log\stet Q,A]\,B\,\big),
\end{equation}
where $[\cdot,\cdot]$ dentes the commutator of operators and $\Res$ stands for the non-commutative residue of Wodzicki
\cite{Wodzicki84,Wodzicki87} and Guillemin \cite{Guillemin85} (see also \cite{Kassel99} for a review).
\eprop
Note, cf. \cite[Prop.~1(i)]{CardonaDucourtiouxMagnotPaycha02}, that $[\log\stet Q,A]$ and, hence, $[\log\stet Q,A]\,B$ are classical \pdo s.
Therefore the non-commutative residue in the right hand side of \refe{notatrace} is well defined.

The following proposition (cf. Lemma~0.1 of \cite{Okikiolu95MA}) describes the dependence of the weighted trace on the operator $Q$.
\prop{TrQ1Q2}
Let $Q_1\in \CLj{m_1},\ Q_2\in \CLj{m_2}$ be positive order classical elliptic \pdo s with Agmon angles $\tet_1$ and $\tet_2$ respectively. For
any $A\in \tCLj{n,\gam}$,
\eq{TrQ1Q2}
    \Tr^{Q_1}\ssub{\tet_1}A \ - \ \Tr^{Q_2}\ssub{\tet_2}A \ = \ \Res\, W,
\end{equation}
where
\eq{W}
    W \ := \ \left(\, \frac{\gam\,\log\ssub{\tet_1}Q_1}{2m_1}\, +\, \frac{\gam\,\log\ssub{\tet_2}Q_2}{2m_2}\,-\,A\,\right)\cdot
    \left(\,  \frac{\log\ssub{\tet_1}Q_1}{m_1}\, -\, \frac{\log\ssub{\tet_2}Q_2}{m_2}\,\right).
\end{equation}
\eprop
It follows immediately from \refe{symlogA} that the operators
\[
    \frac{\gam\,\log\ssub{\tet_1}Q_1}{2m_1}\, +\, \frac{\gam\,\log\ssub{\tet_2}Q_2}{2m_2}\,-\,A \qquad\text{and}\qquad
    \frac{\log\ssub{\tet_1}Q_1}{m_1}\, -\, \frac{\log\ssub{\tet_2}Q_2}{m_2}
\]
are classical. Hence, so is $W$. Thus the non-commutative residue in the right hand side of \refe{TrQ1Q2} is well defined.

Note that if $A\in \CLj{n}$ is a classical \pdo, then $\gam=0$ and \refe{TrQ1Q2} reduces to
\eq{TrQ1Q2classical}
    \Tr^{Q_1}\ssub{\tet_1}A \ - \ \Tr^{Q_2}\ssub{\tet_2}A \ = \ -\,\Res\,
      \left[\,A\cdot\left(\,  \frac{\log\ssub{\tet_1}Q_1}{m_1}\, -\, \frac{\log\ssub{\tet_2}Q_2}{m_2}\,\right)\,\right].
\end{equation}

\subsection{The symmetrized trace}\label{SS:symtrace}
Let $Q\in \CLmj{m}$, where $m$ is a positive integer, and suppose that both, $\tet$ and $\tet-m\pi$, are Agmon angles for $Q$. For $A\in \tCLm$
consider the symmetrized $Q$-trace
\eq{TrsymQ}
    \Tr^{Q,\sym}\stet\,A \ := \ \frac12\,\left(\,\Tr^Q\stet\,A\,+\,\Tr^Q\stetmpi\,A\,\right).
\end{equation}
\prop{Trsym}
Let $Q_j\in \CLmj{m_j}$ $(j=1,2)$ be odd class classical elliptic \pdo s and suppose that $\tet_j$ and $\tet_j-m_j\pi$ are Agmon angles for $Q_j$ ($j=1,2$). Then, for
any odd class logarithmic operator $A\in \tCLmj{n,\gam}$,
\eq{Trsym}
    \Tr^{Q_1,\sym}\ssub{\tet_1}\,A \ = \ \Tr^{Q_2,\sym}\ssub{\tet_2}\,A.
\end{equation}
\eprop
\prf
Set
\[
  \begin{aligned}
    S_{\tet_1,\tet_2} \ &:= \ \frac{\gam\,\log\ssub{\tet_1}Q_1}{2m_1}\ +\ \frac{\gam\,\log\ssub{\tet_2}Q_2}{2m_2} \ - \ A,\\
    T_{\tet_1,\tet_2} \ &:= \ \frac{\log\ssub{\tet_1}Q_1}{m_1}\, -\, \frac{\log\ssub{\tet_2}Q_2}{m_2}.
  \end{aligned}
\]

By \refp{TrQ1Q2},
\eq{TrQ1-TrQ2}
    \Tr^{Q_1,\sym}\ssub{\tet_1}\,A \ - \ \Tr^{Q_2,\sym}\ssub{\tet_2}\,A \ = \ \frac12\,
    \Res\,\big(\,S_{\tet_1,\tet_2}\cdot T_{\tet_1,\tet_2} \,+\,S_{\tet_1-m_1\pi,\tet_2-m_2\pi}\cdot T_{\tet_1-m_1\pi,\tet_2-m_2\pi}\,\big)
\end{equation}
From \refl{logAoddeven} we conclude that $(S_{\tet_1,\tet_2},S_{\tet_1-m_1\pi,\tet_2-m_2\pi}+i\gam\pi)$ and $(T_{\tet_1,\tet_2},T_{\tet_1-m_1\pi,\tet_2-m_2\pi})$ are
odd pairs of operators. Hence, by \refl{oddtimesodd}, the pair
\[
    \big(\,S_{\tet_1,\tet_2}\cdot T_{\tet_1,\tet_2},\,S_{\tet_1-m_1\pi,\tet_2-m_2\pi}\cdot T_{\tet_1-m_1\pi,\tet_2-m_2\pi}+i\gam\pi\, T_{\tet_1-m_1\pi,\tet_2-m_2\pi}\,\big)
\]
is also odd. Therefore,
\[
    S_{\tet_1,\tet_2}\cdot T_{\tet_1,\tet_2}\ + \ S_{\tet_1-m_1\pi,\tet_2-m_2\pi}\cdot T_{\tet_1-m_1\pi,\tet_2-m_2\pi}\ + \ i\gam\pi\, T_{\tet_1-m_1\pi,\tet_2-m_2\pi}
\]
is an odd class classical \pdo. Since the Wodzicki residue of an odd class operator on an odd dimensional manifold vanishes, cf.
\cite[Lemma~7.3]{KontsevichVishik_short}, \cite[Remark~4.5]{KontsevichVishik_long}, it follows that
\eq{=ResT}
    \Res\,\big(\,S_{\tet_1,\tet_2}\cdot T_{\tet_1,\tet_2} \,+\,S_{\tet_1-m_1\pi,\tet_2-m_2\pi}\cdot T_{\tet_1-m_1\pi,\tet_2-m_2\pi}\,\big) \ = \
    -i\gam\pi\, \Res\, T_{\tet_1-m_1\pi,\tet_2-m_2\pi}.
\end{equation}
By \refp{reslog-log}, $\Res T_{\tet_1-m_1\pi,\tet_2-m_2\pi}=0$. Hence \refe{Trsym} follows  from \refe{=ResT} and  \refe{TrQ1-TrQ2}
\eprf

\refp{Trsym} justifies the following
\defe{Trsym}
Let\/ $A\in \tCLm$ be an odd class logarithmic operator on an odd-dimensional closed manifold $M$. The {\em symmetrized trace} $\Tr^{\sym}{A}$
of $A$ is defined by the formula
\eq{Trsymdef}\notag
    \Tr^{\sym} A \ := \ \Tr^{Q,\sym}\stet A,
\end{equation}
where $Q\in \CLmj{m}$ is any odd class classical elliptic \pdo\ of positive order $m$, which admits an Agmon angle, and $\tet\in [0,2\pi)$ is an arbitrary Agmon angle
for $Q$ such that $\tet-m\pi$ is also an Agmon angle for $Q$.
\edefe
The main property of the symmetrized trace is that it is tracial, i.e., vanishes on commutators. More precisely, the following theorem holds.
\th{tracial}
Suppose $M$ is an odd-dimensional closed manifold and let $E$ be a vector bundle over $M$.  Then
\eq{tracial}
    \Tr^{\sym} [A,B] \ = \ 0
\end{equation}
for any operators $A, B\in \tCLm$.
\eth
\prf
Let\/ $Q\in \CLmj{m}$ be an odd class classical elliptic \pdo\ of a positive order $m$, and assume that both, $\tet$ and $\tet-m\pi$, are Agmon angles for $Q$.

Consider first the case when both, $A$ and $B$, are classical. Let  $\tet$ and $\tet-m\pi$ be Agmon angles for $Q$. By \refp{notatracre},
\eq{Tr[AB]}
     \Tr^{Q,\sym} [A,B] \ = \ -\,\frac1{2m}\,\Res\,\Big(\,\big[\,\log\stet Q+\log\stetmpi Q,\,A\,\big]\,B\,\Big).
\end{equation}
By \refc{logAoddodd}, $\log\stet Q+\log\stetmpi Q\in \tCLm$. Hence, by \refl{[logA,B]}(ii),
\[
    \big[\,\log\stet Q+\log\stetmpi Q,\,A\,\big]\,B \ \in \ \CLm.
\]
Since the Wodzicki residue of an odd class operator on an odd dimensional manifold vanishes, cf. \cite[Lemma~7.3]{KontsevichVishik_short},
\cite[Remark~4.5]{KontsevichVishik_long}, equality \refe{tracial} for the case $A,B\in \CLm$ follows from \refe{Tr[AB]}.

Assume now that $A\in \tCLmj{m_1,\gam_1}$, $B\in \tCLmj{m_2,\gam_2}$. Then the operators $A-\frac{\gam_1}{m}\log\stet Q$ and $B-\frac{\gam_2}m
\log\stet Q$ are classical. Hence, it follows from the equality
\eq{[A,B]Q}\notag
    [A,B] \ = \ \big[\, A-\frac{\gam_1}{m}\,\log\stet Q,\,B-\frac{\gam_2}m \,\log\stet Q\,\big] \ + \ \big[\, \log\stet
    Q,\,\frac{\gam_2}m\,A-\frac{\gam_1}m\, B\,\big]
\end{equation}
that
\eq{Tr[A,B]=}
    \Tr^{\sym} [A,B] \ = \ \Tr^{\sym} \big[\, \log\stet Q,\,\frac{\gam_2}m\,A-\frac{\gam_1}m\, B\,\big].
\end{equation}
If $\tet'$ is any Agmon angle for $Q$, then the operators $Q\ssub{\tet'}^s$ and $\log\stet{}Q$ commute. Therefore, for every $s\in
\CC\backslash\ZZ$,
\eq{TRQslogQ}\notag
    \TR Q^s\ssub{\tet'}\big[\, \log\stet Q,\,\frac{\gam_2}m\,A-\frac{\gam_1}m\, B\,\big]
    \ = \ \TR \big[\, Q^s\ssub{\tet'}\log\stet Q,\,\frac{\gam_2}m\,A-\frac{\gam_1}m\, B\,\big] \ = \ 0.
\end{equation}
Hence,
\eq{TrQlogQ}\notag
    \Tr^Q\ssub{\tet'} \big[\, \log\stet Q,\,\frac{\gam_2}m\,A-\frac{\gam_1}m\, B\,\big] \ = \ \Fp \TR Q^s\ssub{\tet'}\big[\, \log\stet Q,\,\frac{\gam_2}m\,A-\frac{\gam_1}m\,
    B\,\big]\ = \ 0.
\end{equation}
From the definition \refe{TrsymQ} of the symmetrized trace we now conclude that
\[
    \Tr^{\sym}\big[\, \log\stet Q,\,\frac{\gam_2}m\,A-\frac{\gam_1}m\, B\,\big]\ =\ 0,
\]
which in view of \refe{Tr[A,B]=} implies \refe{tracial}.
\eprf

\subsection{The Kontsevich-Vishik canonical trace on the algebra of odd class classical \pdo s}\label{SS:canonicaltrace}
For the case when $A\in \CLm$ is an odd class classical \pdo\ Kontsevich and Vishik \cite[\S4]{KontsevichVishik_long} defined a canonical trace $\Tr\ssub{-1}A$ as
follows: Consider a classical elliptic positive definite operator $Q$. Then, cf. \cite[Prop.~4.1]{KontsevichVishik_long}, the function $s\mapsto \TR\,A\,Q^s_{(\pi)}$
is regular at 0 and
\eq{canonicaltrace3}
    \Tr_{(-1)}A \ := \ \TR\,A\,Q^s_{(\pi)}\big|_{s=0}
\end{equation}
is independent of the choice of $Q$. Note that such a $Q$ must have an even order. If we assume that $Q$ is an odd class operator, then by \refe{TrsymQ} we obtain
\eq{canonicaltrace}
     \Tr_{(-1)}A \ := \ \Tr^{Q,\sym}\ssub{\pi}A \ = \ \Tr^\sym A.
\end{equation}

Thus, in the case when $A$ is classical, our symmetrized trace coincides with $\Tr\ssub{-1}$. The advantage of formula \refe{canonicaltrace} is that it allows to
compute the canonical trace using an operator $Q$ which is not positive definite. For example, using the operator $Q$ whose order is odd. This will be important for
applications to determinants in the next section.

\section{The symmetrized determinant}\label{S:symdet}

In this section we define the {\em symmetrized determinant} of an odd class elliptic operator on an odd-dimensional manifold $M$. For operators of even order the
symmetrized determinant coincides with the usual $\zet$-regularized determinant, but for operators of odd order it might be quite different. In the next section we
show that up to sign the symmetrized determinant is very often multiplicative.  This result generalizes the theorem of Kontsevich and Vishik
\cite[Th.~7.1]{KontsevichVishik_short}, \cite[Th.~4.1]{KontsevichVishik_long}, that on odd dimensional manifold the $\zet$-regularized determinant is multiplicative
when restricted to odd class operators, which are close to positive definite ones.

\subsection{Definition of the symmetrized determinant}\label{SS:symdet}
Recall, cf., for example, \cite[\S5]{Ducourtioux99}, that the $\zet$-regularized determinant of an elliptic classical \pdo\ $A\in \CL$ with an
Agmon angle $\tet$ can be defined by the formula
\eq{zetdet}
    \LD\stet A \ = \ \Tr^A\stet \log\stet A \ = \ \frac{d}{ds}\Big|_{s=0}\TR\, A^{s}\stet.
\end{equation}
Note that \refe{zetdet} defines a particular branch of the logarithm of $\Det\stet A$.

We now define a symmetrized version of the $\zet$-regularized determinant as follows:

\defe{symdet}
Let $M$ be an odd dimensional closed manifold and let $E$ be a vector bundle over $M$.  Suppose $A\in \CLmj{m}$ is an odd class classical elliptic \pdo\ of positive
order $m$ and let both, $\tet$ and $\tet-m\pi$, be Agmon angles for $A$. The logarithm of the {\em symmetrized determinant} $\Det^\sym\stet{A}$ is defined by the
formula
\footnote{By \refc{logAoddodd}, $\log\stet A+\log\stetmpi A$ is an odd class logarithmic \pdo. Hence, the symmetrized trace in the right hand side of \refe{symdet2}
is well defined.}
\eq{symdet}
        \LD^\sym\stet A \ := \ \frac12\,\Tr^{\sym}\,\big(\log\stet A+\log\stetmpi A\big).
\end{equation}
\edefe

We now give an alternative formula for the symmetrized determinant.

\prop{symdet}
Under the assumptions of \refd{symdet} we have
\eq{symdet2}
    \LD^\sym\stet A \ := \ \frac12\,\Big(\, \LD\stet A \ + \ \LD\stetmpi A\,\Big).
\end{equation}
\eprop
\prf
Clearly, \refe{symdet2} is equivalent to
\[
    \Tr^A\stet \log\stetmpi A\ + \ \Tr^A\stetmpi\log\stet A \ = \
    \Tr^A\stet\log\stet A \ + \ \Tr^A\stetmpi\log\stetmpi A,
\]
and, hence, to
\eq{Trtet-Trtetmpi}
    \Tr^A\stet\,\big[\, \log\stet A-\log\stetmpi A\,\big] \ - \ \Tr^A\stetmpi\,\big[\, \log\stet A-\log\stetmpi A\,\big] \ = \ 0.
\end{equation}
By \refe{loglogtet1tet2},
\eq{logAtet-logAtmpieven}
    \log\stet A \ - \log\stetmpi A \ = \ 2ik\pi\,\Id,
\end{equation}
if $m=2k$ is even, and
\eq{logAtet-logAtmpiodd}
    \log\stet A\ -\ \log\stetmpi A \ = \ 2ik\pi\,\Id \ + \ 2i\pi\, \Pi_{\tet-\pi,\tet},
\end{equation}
if $m=2k+1$ is odd.  In both cases, $\log\stet A- \log\stetmpi{}A$ is a classical \pdo. Hence, it follows from \refe{TrQ1Q2classical} that the left hand side of
\refe{Trtet-Trtetmpi} is equal to
\eq{Trtet-Trtetmpi2}
    -\frac1m\,\Res\,\Big[\,\big(\,\log\stet A-\log\stetmpi A\,\big)^2\,\Big].
\end{equation}
Since the Wodzicki residue of a pseudo-differential projection vanishes, cf. \cite[\S6]{Wodzicki84}, \cite{Wodzicki87}, we have
\[
    \Res\,\Id \ = \ \Res\,\Pi_{\tet-\pi,\tet} \ = \ 0.
\]
From \refe{logAtet-logAtmpieven} and \refe{logAtet-logAtmpiodd} we conclude now that \refe{Trtet-Trtetmpi2} is equal to 0 and, hence,
\refe{Trtet-Trtetmpi} holds.
\eprf
\prop{detsymeven}
If\/ $m=2k$ is even then $\Det^\sym\stet{}A= \Det\stet{}A$.
\eprop
\prf
As $m$ is even,
\[
    A\stetmpi^s\ =\ e^{-i m\pi s}\,A\stet^s.
\]
Hence, form \refe{zetdet} we obtain
\[
   \LD\stetmpi A \ = \ \LD\stet A \ -\ im\pi\,\TR\,A\stet^s\big|_{s=0}.
\]
Since the $\zet$-function of an odd class elliptic operator of even order on odd dimensional manifold vanishes at 0, cf. \cite{Seeley67}, \cite[\S
II.13]{ShubinPDObook}, \cite[Th.~1.1]{BFK3}, we conclude that $\LD\stetmpi A  = \LD\stet A$. The proposition follows now from \refe{symdet2}.
\eprf

\prop{detsymodd}
If\/ $m=2k+1$ is odd then
\eq{detA2}
    \Det^\sym\ssub{2\tet}(A^2)\ = \ \big(\,\Det^\sym_{(\tet)}A\,\big)^2.
\end{equation}
\eprop
\prf
Let $\Pi_+$ and $\Pi_-$ denote the spectral projections of $A$ corresponding to the solid angles $\Lam_{(\tet-\pi,\tet)}$ and
$\Lam_{(\tet-2\pi,\tet-\pi)}$ respectively. Then $\Pi_+ +  \Pi_-= \Id$. Clearly,
\begin{gather}
     \TR\, (A^2)^{s/2}_{(2\tet)} \  = \ \TR\,\big[\, \Pi_+\,A\stet^{s}\,\big] \ + \  \TR\, \big[\Pi_-\,(-A)\stet^{s}\,\big];\notag\\
 \begin{aligned}
    \TR\, A^s\stet \  &= \ \TR\,\big[\, \Pi_+\,A\stet^{s}\,\big] \ + \ e^{-i\pi s}\, \TR\, \big[\Pi_-\,(-A)\stet^{s}\,\big]
                \\  &= \ \TR\, (A^2)^{s/2}_{(2\tet)} \ + \ (e^{-i\pi s}-1)\, \TR\, \big[\Pi_-\,(-A)\stet^{s}\,\big];
 \end{aligned}\notag
\end{gather}
and
\[
 \begin{aligned}
    \TR\, A^s\stetmpi \  &= \ e^{-i\pi(m+1)s}\,\TR\,\big[\, \Pi_+\,A\stet^{s}\,\big]
                          \ + \ e^{-i\pi m s}\, \TR\, \big[\Pi_-\,(-A)\stet^{s}\,\big]
                \\  &= \ e^{-i\pi(m+1)s}\,\TR\, (A^2)^{s/2}_{(2\tet)}
                        \ + \ (e^{-i\pi m s}-e^{-i\pi(m+1)s})\, \TR\, \big[\Pi_-\,(-A)\stet^{s}\,\big].
 \end{aligned}
\]
Since the Wodzicki residue of a pseudo-differential projection vanishes, cf. \cite[\S6]{Wodzicki84}, \cite{Wodzicki87}, the functions
$\TR\,\big[\, \Pi_+\,A\stet^{s}\,\big]$ and $\TR\, \big[\Pi_-\,(-A)\stet^{s}\,\big]$ are regular at 0. Hence,
\eq{zetprimeAPi+}
    \frac{d}{ds}\big|_{s=0}\,\TR\, A^s\stet
          \ = \ \frac{d}{ds}\big|_{s=0}\,\TR\, (A^2)^{s/2}_{(2\tet)} \ - \ i\pi\,\TR\, \big[\Pi_-\,(-A)\stet^{s}]\big|_{s=0}.
\end{equation}
Similarly,
\meq{zetprimeAPi-}
    \frac{d}{ds}\big|_{s=0}\,\TR\, A^s\stetmpi
          \ = \ \frac{d}{ds}\big|_{s=0}\,\TR\, (A^2)^{s/2}_{(2\tet)}
          \\ -\ i(m+1)\pi\,\TR\, (A^2)^{s/2}_{(2\tet)}\big|_{s=0}  \ + \ i\pi\,\TR\, \big[\Pi_-\,(-A)\stet^{s}]\big|_{s=0}.
\end{multline}
Since the $\zet$-function of an odd class elliptic operator of even order on odd dimensional manifold vanishes at 0, cf. \cite{Seeley67},
\cite[\S II.13]{ShubinPDObook}, \cite[Th.~1.1]{BFK3}, we conclude from \refe{zetprimeAPi-} that
\eq{zetprimeAPi-2}
    \frac{d}{ds}\big|_{s=0}\,\TR\, A^s\stetmpi
          \ = \ \frac{d}{ds}\big|_{s=0}\,\TR\, (A^2)^{s/2}_{(2\tet)} \ + \ i\pi\,\TR\, \big[\Pi_-\,(-A)\stet^{s}]\big|_{s=0}.
\end{equation}
From \refe{zetprimeAPi+}, \refe{zetprimeAPi-2} and \refe{symdet2} we obtain \refe{detA2}.
\eprf

\subsection{Dependence on the angle}\label{SS:deponangle}
For the usual $\zet$-regularized determinant $\Det\stet{}A$, we have $\Det\stet A= \Det\ssub{\tet'}A$ if there are only finitely many
eigenvalues of $A$ in the solid angle
\[
    \Lam_{(\tet,\tet')} \ := \ \big\{ \rho\,e^{i\alp}:\ \rho\ge0,\, \tet\le\alp\le\tet'\,\big\}.
\]
If the order $m$ of $A$ is even, then in view of \refp{detsymeven} the same is true for the symmetrized determinant $\Det^\sym\stet{}A$.
However, if $m$ is odd, then the above condition does not imply that $\Det^\sym\stet A  = \Det^\sym\ssub{\tet'}A$ since it might happen that
there are infinitely many eigenvalues of $A$ in the solid angle $\Lam_{(\tet-\pi,\tet'-\pi)}$. Moreover, because of the factor $1/2$ in the
definition \refe{symdet} even if there are only finitely many eigenvalues of $A$ in the solid angles $\Lam_{(\tet,\tet')}$ and
$\Lam_{(\tet-\pi,\tet'-\pi)}$, we only have
\eq{fneigenv}
 \Det^\sym\stet A \ =\ \pm\,\Det^\sym\ssub{\tet'}A.
\end{equation}
If the are infinitely many eigenvalues in $\Lam_{(\tet,\tet')}$, then \refe{fneigenv} does not hold in general. However, the dependence of
$\Det\stet{}A$ on $\tet$ in this case is weaker than the dependence of the usual $\zet$-regularized determinant. In particular, the following
proposition holds.

\prop{deponangle}
Let $M$ be an odd-dimensional closed manifold and let $E$ be a vector bundle over $M$. Suppose $A\in \CLmj{m}$ is an elliptic \pdo\ of odd
positive  order $m=2l-1$ and that $\tet_i$ and $\tet_i-\pi$ $(i=1,2)$ are Agmon angles for $A$. Suppose that $0\le \tet_2-\tet_1<\pi$ and that
all but finitely many eigenvalues of $A$ lie in $\Lam_{(\tet_1,\tet_2)}\cup\Lam_{(\tet_1-\pi,\tet_2-\pi)}$. Then
\eq{deponangle}
    \Det_{(\tet_1)}^\sym A \ = \ \pm\, \Det_{(\tet_2)}^\sym A.
\end{equation}
\eprop

\prf
Since $m$ is odd it follows from our assumptions that $\tet_1-m\pi$ and $\tet_2-m\pi$ are Agmon angles for $A$. Let $\Pi_{\tet_1,\tet_2}$ and
$\Pi_{\tet_1-m\pi,\tet_2-m\pi}$ be as in \refss{angle}. Then
\eq{Atet1s-Atet2s}
 \begin{aligned}
    A_{(\tet_1)}^s \ - \ A_{(\tet_2)}^s \ &= \ (1-e^{2\pi i s})\, \Pi_{\tet_1,\tet_2}\,A_{(\tet_1)}^s;\\
    A_{(\tet_1-m\pi)}^s \ - \ A_{(\tet_2-m\pi)}^s \ &= \ (1-e^{2\pi i s})\, \Pi_{\tet_1-m\pi,\tet_2-m\pi}\,A_{(\tet_1-m\pi)}^s.
 \end{aligned}
\end{equation}
Since the Wodzicki residue of a pseudo-differential projection vanishes, cf. \cite[\S6]{Wodzicki84}, \cite{Wodzicki87}, the functions
\[
    s\ \mapsto\  \TR\,\big[\,\Pi_{\tet_1,\tet_2}\,A_{(\tet_1)}^s\,\big] \qquad \text{and}
    \qquad s\ \mapsto\  \TR\,\big[\,\Pi_{\tet_1-m\pi,\tet_2-m\pi}\,A_{(\tet_1-m\pi)}^s\,\big]
\]
are regular at $s=0$.  Hence, from \refe{zetdet}, \refe{symdet}, and \refe{Atet1s-Atet2s}, we obtain
\eq{symdet=zet}
    \LD_{(\tet_1)}^\sym A \ - \ \LD_{(\tet_2)}^\sym A \ = \
    -\,\pi i\, \TR\Big[\,\Pi_{\tet_1,\tet_2}\,A_{(\tet_1)}^s+ \Pi_{\tet_1-m\pi,\tet_2-m\pi}\,A_{(\tet_1-m\pi)}^s\,\Big]\Big|_{s=0}.
\end{equation}

Let $\tilPi$ denote the spectral projection of the operator $A^2$ corresponding to the eigenvalues of $A^2$ which lie in the solid angle
$\Lam_{2\tet_1,2\tet_2}$. Then
\eq{P=P+P}
    \tilPi\ = \ \Pi_{\tet_1,\tet_2} \ + \ \Pi_{\tet_1-m\pi,\tet_2-m\pi}
\end{equation}
and
\[
    \TR\big[\,\Pi_{\tet_1,\tet_2}\,A_{(\tet_1)}^s\,\big]\ + \ e^{m\pi i s}\,\TR\,\big[\, \Pi_{\tet_1-m\pi,\tet_2-m\pi}\,A_{(\tet_1-m\pi)}^s\,\big]
    \ = \  \TR\,\big[\,\tilPi\,(A^2)^{s/2}\ssub{2\tet_1}\,\big].
\]
Hence, at $s=0$ we obtain
\eq{A-A2}
     \TR\Big[\,\Pi_{\tet_1,\tet_2}\,A_{(\tet_1)}^s+ \Pi_{\tet_1-m\pi,\tet_2-m\pi}\,A_{(\tet_1-m\pi)}^s\,\Big]\Big|_{s=0} \ = \
      \TR\,\big[\,\tilPi\,(A^2)^{s/2}\ssub{2\tet_1}\,\big]\Big|_{s=0}.
\end{equation}
From \refe{P=P+P} and the assumptions of the proposition we conclude that $I-\tilPi$ is a finite dimensional projection. Hence,
\eq{I-tilPi}\notag
       \TR\,\big[\,(A^2)^{s/2}\ssub{2\tet_1}\,\big]\big|_{s=0} \ - \
      \TR\,\big[\,\tilPi\,(A^2)^{s/2}\ssub{2\tet_1}\,\big]\big|_{s=0}\ \in\ \ZZ.
\end{equation}
Since the $\zet$-function of an odd class elliptic operator of even order on an odd dimensional manifold vanishes at 0, cf. \cite{Seeley67},
\cite[\S{II.13}]{ShubinPDObook}, we have $\TR\big[\,(A^2)^{s/2}\ssub{2\tet_1}\big]\big|_{s=0}=0$, and, hence,
\[
      \TR\,\big[\,\tilPi\,(A^2)^{s/2}\ssub{2\tet_1}\,\big]\big|_{s=0}\ \in\ \ZZ.
\]
Therefore, from \refe{symdet=zet} and \refe{A-A2} we obtain
\eq{LD-LD}
    \LD_{(\tet_1)}^\sym A \ - \ \LD_{(\tet_2)}^\sym A \ \in \ \pi i\,\ZZ,
\end{equation}
which is equivalent to \refe{deponangle}.
\eprf

\cor{deponangle-sa}
Let $A\in \CLmj{m}$ be an elliptic operator of odd order $m=2l-1$ on an odd dimensional manifold $M$. Assume that the leading symbol of $A$ is self-adjoint with
respect to the scalar product defined by some Hermitian metric on $E$. Then up to sign the symmetrized determinant $\Det\stet^\sym{}A$ is independent of the choice of
the Agmon angle $\tet$.
\ecor
\rem{KVdet}
Kontsevich and Vishik, \cite[\S4.1]{KontsevichVishik_long}, suggested to define a determinant of a Dirac-type operator $A$ as a square root of the determinant of the
Laplace-type operator $A^2$. Thus their definition had a sign indeterminacy. It follows from the equation \refe{detA2}  that the reduction modulo signs of the
symmetrized determinant $\Det\ssub{\pi/2}^\sym{}A$ coincides in the case of Dirac-type operators with the Kontsevich-Vishik determinant. Thus in the case of
Dirac-type operators our symmetrized determinant provides a more direct construction of the Kontsevich-Vishik determinant and also fixes the signs in its definition.
\erem

\section{The Multiplicative Properties of the Symmetrized Determinant}\Label{S:multprop}

In this section we show, that under suitable assumptions the symmetrized determinant is multiplicative up to sign. Let $\tet$ be a principal
angle for an elliptic operator $A\in \CLmj{m}$. We say that an Agmon angle $\tet'\ge\tet$ is sufficiently close to $\tet$ if there are no
eigenvalues of $A$ in the solid angles $\Lam_{(\tet,\tet']}$ and $\Lam_{(\tet-m\pi,\tet'-m\pi]}$. We shall denote by $\Det\ssub{\tiltet}A,\
\Det^\sym\ssub{\tiltet}A,\ \log\ssub{\tiltet}A$, etc. the corresponding numbers defined using any Agmon angle $\tet'\ge\tet$ sufficiently closed
to $\tet$. Clearly, those numbers are independent of the choice of such $\tet'$.
\th{multprop}
Let $M$ be an odd dimensional manifold and let $E$ be a vector bundle over $M$. Suppose $A\in \CLmj{m_A}$ and $B\in \CLmj{m_B}$ are odd class
classical invertible \pdo s of positive integer orders. Let\/ $\tet_A$ and $\tet_B$ be principal angles for $A$ and $B$ respectively. Assume
further that there exists a continuous functions $\alp:[0,1]\to \RR$ such that $\alp(0)= \tet_B$ and for each $t\in [0,1]$ the angle $\alp(t)$
is principal for the operator $A_{(\widetilde\tet_A)}^tB$.  Set $\tet_B=\alp(0),\ \tet_{AB}= \alp(1)$. Then
\eq{multprop}
    \Det_{(\widetilde\tet_{AB})}^\sym AB \ = \ \pm\,\Det_{(\widetilde\tet_A)}^\sym A\cdot \Det_{(\widetilde\tet_B)}^\sym B.
\end{equation}
\eth
The proof is given in \refss{prmultprop}.

The assumptions of \reft{multprop} are rather restrictive. There are, however, several important situations, where these conditions are
automatically satisfied. We now mention some of these situations.

\cor{multsapd}
Assume that the leading symbols of $A$ and $B$ are self-adjoint and that the leading symbol of $A$ is positive definite. Then \refe{multprop} holds
with $\tet_{AB}= \tet_B$.
\ecor
\prf
Without loss of generality we can assume that $\tet_B\not\in \pi\ZZ$. The leading symbol of the operator $A^t_{(\widetilde\tet_A)}B$ is
conjugate to the self-adjoint symbol
\[
    \sig_{tm_A/2}\big(\,A_{(\widetilde\tet_A)}^{t/2}BA_{(\widetilde\tet_A)}^{t/2}\,\big).
\]
Thus $\tet_B$ is a principal angle for $A^t_{(\tet_A)}B$. Hence, we can apply \reft{multprop} with $\alp(t)=\tet_B$.
\eprf
\cor{multcom}
Assume that the leading symbols of\/ $A$ and $B$ commute with each other and that the leading symbol of $A$ is self-adjoint. Let $\tet_A,
\tet_B\in (0,2\pi)$ be principal angles for $A$ and $B$, respectively. Then  equality \refe{multprop} holds with\/ $\tet_{AB}=
\tet_A+\eps\tet_B$, where $\eps=1$ if $0<\tet_A<\pi$ and $\eps=-1$ if $\pi<\tet_A<2\pi$.
\ecor
\prf
Since the leading symbol $\sig_{m_A}(A)$ of $A$ is self-adjoint, all its eigenvalues lie on the real axis. Hence, all the eigenvalues of
\[
    \sig_{tm_A}\big(\,A_{(\widetilde\tet_A)}^t\,\big) \ = \ \big(\,\sig_{m_A}(A)\,\big)^t_{(\tet_A)}
\]
lie on the rays $R_0= \{r:\,r>0\}$ and $R_{t\eps\pi}= \{re^{-i\eps\pi}:\,r>0\}$. Thus, since the leading symbols of $A$ and $B$ commute the
eigenvalues of the leading symbol
\[
    \sig_{tm_A+m_B}\big(\,A_{(\widetilde\tet_A)}^tB\,\big) \ =\  \big(\,\sig_{m_A}(A)\,\big)^t_{(\tet_A)}\cdot\sig_{m_B}(B)
\]
have the form $r\lam$, $r\lam{}e^{-i\eps t}$, where $\lam$ is an eigenvalue of $\sig_{m_B}(B)$ and $r>0$. Therefore we can apply \reft{multprop}
with $\alp(t)=  \tet_B+t\eps\tet_A$.
\eprf

\subsection{Proof of \reft{multprop}}\Label{SS:prmultprop}
Without loss of generality we can assume that $\tet_A$ and $\tet_A-m_A\pi$ are Agmon angles for $A$,  and $\tet_B$ and $\tet_B-m_B\pi$ are Agmon
angles for $B$.

The leading symbol of the operator $A^t_{(\tet_A)}B$ is given by
\[
    \sig_{tm_A+m_B}(A^t_{(\tet_A)}B)(x,\xi) \ = \ \big(\,\sig_{m_A}(A)(x,\xi)\,\big)^t_{(\tet_A)}\cdot \sig_{m_B}(B)(x,\xi).
\]
Hence, from \refe{Asoddodd} we obtain
\[
  \begin{aligned}
    \sig_{tm_A+m_B}(A^t_{(\tet_A)}B)(x,-\xi) \ &=  \ (-1)^{m_B}\,e^{im_At\pi}\,\sig_{tm_A+m_B}(A^t_{(\tet_A-m_A\pi)}B)(x,\xi)
                                            \\ &=   \ e^{i\,(m_At+m_B)\,\pi}\,\sig_{tm_A+m_B}(A^t_{(\tet_A-m_A\pi)}B)(x,\xi).
  \end{aligned}
\]
Therefore, if $\alp(t)$ is a principal angle for $A^t_{(\tet_A)}B$ then
\[
     \bet(t) \ := \ \alp(t)\ -\ i\,(m_At+m_B)\,\pi
\]
is a principal angle for $A^t_{(\tet_A-m_A\pi)}B$.

Using the formula for multiplicative anomaly of the $\zet$-regularized determinant, \cite[Prop.~2.1]{KontsevichVishik_long}
\cite[Th.~1]{Okikiolu95MA}, we conclude that
\meq{anomaly}
    \LD^\sym_{(\tet_{AB})}AB \ - \ \LD^\sym_{(\tet_A)}A \ - \ \LD^\sym_{(\tet_B)}B
    \\ \equiv \ \frac{m_A}4\,\int_0^1\Res\,\big(\,U(t)^2 + V(t)^2\,\big)\,dt, \qquad \MOD \ \pi\ZZ,
\end{multline}
where
\eq{W1}
    U(t) \ = \ \frac{\log\ssub{\widetilde\alp(t)}A^t_{(\tet_A)}B}{m_At+m_B}\ - \ \frac{\log_{(\tet_A)}A}{m_A},
\end{equation}
and
\eq{W2}
    V(t) \ = \ \frac{\log\ssub{\widetilde\bet(t)}A^t_{(\tet_A-m_A\pi)}B}{m_At+m_B}\ -\ \frac{\log_{(\tet_A-m_A\pi)}A}{m_A}.
\end{equation}

\lem{U-V}
For each $t\in [0,1]$, $(U(t),V(t))$ is an odd pair of operators in the sense of \refss{oddpair}.
\elem
\prf
Fix $t\in [0,1]$ and set $m=tm_A+m_B$. From \refe{Asoddodd} we easily obtain
\eq{sigAtB}
    \sig_{m-j}\big(A\ssub{\tet_A}^tB\big)(x,-\xi) \ = \ (-1)^j\,e^{im\pi}\,\sig_{m-j}\big(A\ssub{\tet_A-m_A\pi}^tB\big)(x,\xi).
\end{equation}
Set
\[
    R(\lam) \ := \ \big(\, A\ssub{\tet_A}^tB-\lam\,\big)^{-1}, \qquad \hatR(\lam) \ := \ \big(\, A\ssub{\tet_A-m_A\pi}^tB-\lam\,\big)^{-1}.
\]
Let $r(x,\xi;\lam)\sim \sum_{j=0}^\infty r_{-m-j}(x,\xi;\lam)$ and $\hatr(x,\xi;\lam)\sim \sum_{j=0}^\infty \hatr_{-m-j}(x,\xi;\lam)$ be the symbols of $R(\lam)$ and
$\hatR(\lam)$ respectively. Here the asymptotic expansions are understood in the same sense as in \refe{rxxi}.

From \refe{sigAtB} and the standard formulae for the parametrix, cf., for example, \cite[\S2]{KontsevichVishik_long}, we get
\eq{rAtB}
    r_{-m-j}(x,-\xi;e^{im\pi}\lam) \ = \ (-1)^j\,e^{im\pi}\, \hatr_{-m-j}(x,\xi;\lam).
\end{equation}
Hence, a verbatim repetition of the computation in the proof of \refl{Asodd} yields
\eq{AtBs}
    \sig_{sm-j}\big(\,(A_{(\tet_A)}^tB)^s\ssub{\widetilde\alp(t)}\,\big)(x,-\xi)
     \ = \
    (-1)^j\,e^{ism\pi}\, \sig_{sm-j}\big(\,(A_{(\tet_A-m_A\pi)}^tB))^s\ssub{\widetilde\bet(t)}\,\big)(x, \xi).
\end{equation}
Since $(A_{(\tet_A)}^tB)^s\ssub{\widetilde\alp(t)}\big|_{s=0} = (A_{(\tet_A-m_A\pi)}^tB))^s\ssub{\widetilde\bet(t)}\big|_{s=0}= \Id$, we obtain
\eq{ddsAsB}
    \partial_s\,\sig_{sm-j}\big(\,(A_{(\tet_A)}^tB)^s\ssub{\widetilde\alp(t)}\,\big)(x,\xi)\big|_{s=0} \ = \ \del_{j,0}\,\Id.
\end{equation}
Combining \refe{AtBs} with \refe{ddsAsB} we get
\meq{ddsAsBoddd}
  \partial_s\,\sig_{sm-j}\big(\,(A_{(\tet_A)}^tB)^s\ssub{\widetilde\alp(t)}\,\big)(x,-\xi/|\xi|)\big|_{s=0} \\ = \
  i\,m\,\pi\,\del_{j,0} \ + \
  \partial_s\,\sig_{sm-j}\big(\,(A_{(\tet_A-m_A\pi)}^tB))^s\ssub{\widetilde\bet(t)}\,\big)(x, \xi/|\xi|)\big|_{s=0}.
\end{multline}
Hence, using \refe{symlogA=} we conclude that
\[
     \Big(\,\log_{(\widetilde\alp(t))}A^t_{(\tet_A)}B,\, \log_{(\widetilde\bet(t))}A_{(\tet_A-m_A\pi)}^tB+i\,m\,\pi\Id\,\Big)
\]
is an odd pair of operators. The statement of the lemma follows now from \refl{logAoddeven} and the definitions of the operators $U(t),\ V(t)$.
\eprf

Combining \refl{oddtimesodd} and \refl{U-V} we conclude that $(U(t)^2,V(t)^2)$ is an odd pair of operators. Hence, the operator $U(t)^2+V(t)^2$ is of odd class. Since
the Wodzicki residue of an odd class operator on an odd dimensional manifold vanishes, cf. \cite[Lemma~7.3]{KontsevichVishik_short},
\cite[Remark~4.5]{KontsevichVishik_long}, we obtain $\Res\big(\,U(t)^2 + V(t)^2\,\big)=0$. Therefore, \refe{multprop} follows from \refe{anomaly}.\hfill$\square$

\section{Operators whose Spectrum is Symmetric about the Real Axis}\Label{S:det-eta-sa}

In this section we prove that the symmetrized determinant of an operator whose spectrum is symmetric about the real axis is real and we compute its sign. Note that
the spectrum of every differential operator with {\em real coefficients} is symmetric about the real axis. Other interesting examples of operators with symmetric
spectrum are discussed in \cite{AbanovWieg00,Abanov-Hopf,BrAbanov}.

\defe{realcoef}
The spectrum of $A$ is {\em symmetric  with respect to the real axis} if the following condition holds: if\/ $\lam$ is an eigenvalue of $A$,
then $\olam$ also is an eigenvalue of\/ $A$ and has the same algebraic multiplicity as $\lam$.
\edefe

We denote by $P_+$ and $P_-$ the spectral projection of $A$ corresponding to the positive and negative parts of the imaginary axis respectively. Assume that the
leading symbol of $A$ is self-adjoint with respect to some Hermitian metric on $E$. Then the projections $P_\pm$ have finite rank. Set $m_\pm= \rank{}P_\pm$. Then
$m_+$ (respectively $m_-$) is equal to the number of the eigenvalues of $A$ (counted with their algebraic multiplicities) on the positive (respectively negative) part
of the imaginary axis. If the spectrum of $A$ is symmetric about the real axis then $m_+= m_-$.

\th{realcoef}
Let $A\in \CLmj{m}$ be a classical elliptic pseudo-differential operator of positive odd order $m=2l-1$, whose spectrum is symmetric about the real axis and whose
leading symbol is self-adjoint with respect to some Hermitian metric on $E$. Choose $\tet\in (\pi/2,\pi)$ such that both, $\tet$ and $\tet-m\pi$, are Agmon angles for
$A$ and there are no eigenvalues of $A$ in the solid angles $\Lam_{(\pi/2,\tet]}$ and $\Lam_{(-\pi/2,\tet-\pi]}$. Then the symmetrized determinant $\Det^\sym\stet{}A$
is real and its sign is equal to $(-1)^{m_+}$, i.e.
\eq{realcoef}
    \Det^\sym\stet A \ = \ (-1)^{m_+}\,\big|\,\Det^\sym\stet A\,\big|.
\end{equation}
\eth
\prf
Let $\Pi_+$ and $\Pi_-$ denote the spectral projections of $A$ corresponding to the solid angles $L_{(-\pi/2,\pi/2)}$ and $L_{(\pi/2,3\pi/2)}$
respectively. Let $P_+$ and $P_-$ denote the spectral projections of $A$ corresponding to the rays $R_{\pi/2}$ and $R_{-\pi/2}$ respectively
(here we use the notation introduced in \refss{complexpower}). Set $\tilPi_\pm= \Pi_\pm+P_\pm$. Since $A$ is injective
 \(
    \Id \ = \ \tilPi_+ \ + \  \tilPi_-.
 \)
Clearly,
\begin{gather}
    \TR A\stet^s \ = \ \TR\,\big[\, \tilPi_+\,A\stet^s\,\big] \ + \ e^{-i\pi s}\, \TR\, \big[\tilPi_-\,(-A)\stet^s\,\big];\\
    \TR A\stetmpi^s \ = \ e^{-i(m+1)\pi s}\, \TR\,\big[\, \tilPi_+\,A\stet^s\,\big] \ + \
            e^{-im\pi s}\, \TR\, \big[\tilPi_-\,(-A)\stet^s\,\big];\\
    \TR (A^2)^{s/2}\ssub{2\tet} \  = \ \TR\,\big[\, \tilPi_+\,A\stet^{s}\,\big] \ + \  \TR\, \big[\tilPi_-\,(-A)\stet^{s}\,\big].\label{E:A2s}
\end{gather}
Hence,
\meq{TRAstet+TRAstetpi}
    \TR A\stet^s \ + \ \TR A\stetmpi^s \\ = \ (1+e^{-i(m+1)\pi s})\,\TR\,\big[\, \tilPi_+\,A\stet^s\,\big]
       \ + \ (e^{-i\pi s}+e^{-im\pi s})\, \TR\, \big[\tilPi_-\,(-A)\stet^s\,\big].
\end{multline}
Since the functions $\TR\,\big[\, \tilPi_+\,A\stet^s\,\big]$ and $\TR\, \big[\tilPi_-\,(-A)\stet^s\,\big]$ are regular at $s=0$ we conclude from
\refe{TRAstet+TRAstetpi} that
\meq{ddstr+tr}
    \frac{d}{ds}\Big|_{s=0}\,\TR A\stet^s\ + \ \frac{d}{ds}\Big|_{s=0}\,\TR A\stetmpi^s \\ = \
    2\,\frac{d}{ds}\Big|_{s=0}\,\TR\,\big[\, \tilPi_+\,A\stet^s\,\big]\ + \ 2\,\frac{d}{ds}\Big|_{s=0}\,\TR\, \big[\tilPi_-\,(-A)\stet^s\,\big]
    \\ - \ i\,(m+1)\,\pi\,\TR\,\big[\, \tilPi_+\,A\stet^s\,\big] \ - \ \big(\,i\,\pi\,+\,i\,m\,\pi\,\big)\,\TR\, \big[\tilPi_-\,(-A)\stet^s\,\big]
    \\ = \
    2\,\frac{d}{ds}\Big|_{s=0}\,\TR\,\big[\, \tilPi_+\,A\stet^s\,\big]\ + \ 2\,\frac{d}{ds}\Big|_{s=0}\, \TR\, \big[\tilPi_-\,(-A)\stet^s\,\big]
    \\ - \ i\,(m+1)\,\pi\,\Big[\,\TR\,\big[\, \tilPi_+\,A\stet^{s}\,\big] \ + \  \TR\, \big[\tilPi_-\,(-A)\stet^{s}\,\big]\,\Big]\Big|_{s=0}.
\end{multline}
Using \refe{A2s} and the fact that the $\zet$-function of an odd class elliptic operator of even order on odd dimensional manifold vanishes at 0, cf. \cite{Seeley67},
\cite[\S II.13]{ShubinPDObook}, \cite[Th.~1.1]{BFK3}, we conclude that
\[
    \Big[\,\TR\,\big[\, \tilPi_+\,A\stet^{s}\,\big] \ + \  \TR\, \big[\tilPi_-\,(-A)\stet^{s}\,\big]\,\Big]\Big|_{s=0} \ = \ 0.
\]
Also by formula (A.2) of \cite{BrKappelerRAT} that the imaginary part of
\[
    \frac{d}{ds}\Big|_{s=0}\,\TR\,\big[\, \tilPi_+\,A\stet^s\,\big]\ + \ \frac{d}{ds}\Big|_{s=0}\, \TR\, \big[\tilPi_-\,(-A)\stet^s\,\big]
\]
is equal to $-m_+\pi$. Hence, we conclude from \refe{ddstr+tr} that the imaginary part of
\[
    \LD\stet^\sym A \ = \ \frac12\, \Big[\,\frac{d}{ds}\Big|_{s=0}\,\TR A\stet^s\ + \ \frac{d}{ds}\Big|_{s=0}\,\TR A\stetmpi^s \,\Big]
\]
is equal to $-m_+\pi$.
\eprf

\providecommand{\bysame}{\leavevmode\hbox to3em{\hrulefill}\thinspace} \providecommand{\MR}{\relax\ifhmode\unskip\space\fi MR }
\providecommand{\MRhref}[2]{%
  \href{http://www.ams.org/mathscinet-getitem?mr=#1}{#2}
} \providecommand{\href}[2]{#2}


\begin{thebibliography}{10}

\bibitem{BrAbanov}
A.~Abanov and M.~Braverman, \emph{{Topological calculation of the phase of the
  determinant of a non self-adjoint elliptic operator}}, Comm. Math. Phys.
  \textbf{259} (2005), 287--305.

\bibitem{Abanov-Hopf}
A.~G. Abanov, \emph{Hopf term induced by fermions}, Phys.Lett. \textbf{B492}
  (2000), 321--323.

\bibitem{AbWie-geom}
A.~G. Abanov and P.~B. Wiegmann, \emph{Geometrical phases and quantum numbers
  of solitons in nonlinear sigma-models}, Preprint, \texttt{hep-th/0105213 }.

\bibitem{AbanovWieg00}
\bysame, \emph{Theta-terms in nonlinear sigma-models}, Nucl.Phys. \textbf{B570}
  (2000), 685--698.

\bibitem{AbWie-chir}
\bysame, \emph{Chiral non-linear sigma-models as models for topological
  superconductivity}, Phys.Rev.Lett. \textbf{86} (2001), 1319--1322.

\bibitem{BrKappelerRAT}
M.~Braverman and T.~Kappeler, \emph{{Refined Analytic Torsion}},
  \texttt{arXiv:math.DG/0505537}, {\em To appear in} Journal of Differential Geometry.

\bibitem{BFK3}
D.~Burghelea, L.~Friedlander, and T.~Kappeler, \emph{Asymptotic expansion of
  the {Witten} deformation of the analytic torsion}, Journal of Funct. Anal.
  \textbf{137} (1996), 320--363.

\bibitem{CardonaDucourtiouxMagnotPaycha02}
A.~Cardona, C.~Ducourtioux, J.~P. Magnot, and S.~Paycha, \emph{Weighted traces
  on algebras of pseudo-differential operators and geometry on loop groups},
  Infin. Dimens. Anal. Quantum Probab. Relat. Top. \textbf{5} (2002), no.~4,
  503--540.

\bibitem{Ducourtioux99}
C.~Ducourtioux, \emph{Multiplicative anomaly for the {$\zeta$}-regularized
  determinant}, Geometric methods for quantum field theory (Villa de Leyva,
  1999), World Sci. Publishing, River Edge, NJ, 2001, pp.~467--482.

\bibitem{Guillemin85}
V.~Guillemin, \emph{A new proof of {W}eyl's formula on the asymptotic
  distribution of eigenvalues}, Adv. in Math. \textbf{55} (1985), 131--160.

\bibitem{Kassel99}
C.~Kassel, \emph{Le r\'esidu non commutatif (d'apr\`es {M}.\ {W}odzicki)},
  Ast\'erisque (1989), no.~177-178, Exp.\ No.\ 708, 199--229, S\'eminaire
  Bourbaki, Vol.\ 1988/89.

\bibitem{KontsevichVishik_long}
M.~Kontsevich and S.~Vishik, \emph{{Determinants of elliptic
  pseudo-differential operators}}, Preprint MPI /94-30,
  \texttt{arXiv:hep-th/9404046}.

\bibitem{KontsevichVishik_short}
\bysame, \emph{Geometry of determinants of elliptic operators}, Functional
  analysis on the eve of the 21st century, Vol.\ 1 (New Brunswick, NJ, 1993),
  Progr. Math., vol. 131, Birkh\"auser Boston, Boston, MA, 1995, pp.~173--197.

\bibitem{Lesch99}
M.~Lesch, \emph{On the noncommutative residue for pseudodifferential operators
  with log-polyhomogeneous symbols}, Ann. Global Anal. Geom. \textbf{17}
  (1999), no.~2, 151--187.

\bibitem{MelroseNistor96}
R.~B. Melrose and V.~Nistor, \emph{{Homology of pseudodifferential operators I.
  Manifolds with boundary}}, \texttt{arXiv:funct-an/9606005}.

\bibitem{Okikiolu95CH}
K.~Okikiolu, \emph{The {C}ampbell-{H}ausdorff theorem for elliptic operators
  and a related trace formula}, Duke Math. J. \textbf{79} (1995), no.~3,
  687--722.

\bibitem{Okikiolu95MA}
\bysame, \emph{The multiplicative anomaly for determinants of elliptic
  operators}, Duke Math. J. \textbf{79} (1995), no.~3, 723--750.

\bibitem{Redlich84}
A.~N. Redlich, \emph{Gauge noninvariance and parity nonconservation of
  three-dimensional fermions}, Phys. Rev. Lett. \textbf{52} (1984), 18--21.

\bibitem{Redlich84a}
\bysame, \emph{Parity violation and gauge noninvariance of the effective gauge
  field action in three dimensions}, Phys. Rev. D (3) \textbf{29} (1984).

\bibitem{Seeley67}
R.~Seeley, \emph{Complex powers of elliptic operators}, Proc. Symp. Pure and
  Appl. Math. AMS \textbf{10} (1967), 288--307.

\bibitem{ShubinPDObook}
M.~A. Shubin, \emph{Pseudodifferential operators and spectral theory}, Springer
  Verlag, Berlin, New York, 1980.

\bibitem{Wodzicki84}
M.~Wodzicki, \emph{Local invariants of spectral asymmetry}, Invent. Math.
  \textbf{75} (1984), no.~1, 143--177.

\bibitem{Wodzicki87}
\bysame, \emph{Noncommutative residue. {I}. {F}undamentals}, $K$-theory,
  arithmetic and geometry (Moscow, 1984--1986), Lecture Notes in Math., vol.
  1289, Springer, Berlin, 1987, pp.~320--399.

\end{thebibliography}
\end{document}